\documentclass[preprint, a4paper, 12pt, 3p]{elsarticle}
\usepackage[british]{babel}
\usepackage{amssymb,amsmath,amsfonts}
\usepackage{graphicx}
\usepackage{subfigure}
\def\vet#1{{\underline #1}}
\def\build#1_#2^#3{\mathrel{
\mathop{\kern 0pt#1}\limits_{#2}^{#3}}}
\def\reali{\mathbb{R}}
\def\naturali{\mathbb{N}}
\def\interi{\mathbb{Z}}

\def\Dscr{\mathcal{D}}
\def\Fscr{\mathcal{F}}
\def\Gscr{\mathcal{G}}
\def\Hscr{\mathcal{H}}

\def\Lscr{\mathcal{L}}
\def\Oscr{\mathcal{O}}

\def\Tscr{\mathcal{T}}
\def\lie#1{\Lscr_{#1}}
\def\gt{\ge}
\def\lt{\le}
\def\epsilon{\varepsilon}
\def\rho{\varrho}
\def\csi{\xi}

\def\adaptpoisson#1#2{\left\{ #1\,,\,#2 \right\}}
\def\fastpoisson#1#2{\left\{ #1,#2 \right\}_{\vet{L},\vet{\lambda}}}
\def\secpoisson#1#2{\left\{ #1,#2 \right\}_{\vet{\csi},\vet{\eta}}}
\begin{document}

\markboth{M. Sansottera, U. Locatelli, A. Giorgilli}{On the stability
  of the secular $\ldots$ planar Sun--Jupiter--Saturn--Uranus system}

\begin{frontmatter}

\title{On the stability of the secular evolution of the
  planar Sun--Jupiter--Saturn--Uranus system}

\author[mi]{M.~Sansottera}
\ead{marco.sansottera@gmail.com}

\author[ro]{U.~Locatelli}
\ead{locatell@mat.uniroma2.it}

\author[mi]{A.~Giorgilli}
\ead{antonio.giorgilli@unimi.it}

\address[mi]{Dipartimento di Matematica, Universit\`a degli Studi di Milano,\\
  via Saldini 50, 20133\ ---\ Milano, Italy.}
\address[ro]{Dipartimento di Matematica, Universit\`a degli Studi di Roma ``Tor Vergata'',\\
  via della Ricerca Scientifica 1, 00133\ ---\ Roma, Italy.}

\begin{abstract}
We investigate the long time stability of the
Sun--Jupiter--Saturn--Uranus system by considering the planar, secular
model.  Our method may be considered as an extension of Lagrange
theory for the secular motions.  Indeed, concerning the planetary
orbital revolutions, we improve the classical circular approximation
by replacing it with a torus which is invariant up to order two in the
masses; therefore, we investigate the stability of the elliptic
equilibrium point of the secular system for small values of the
eccentricities.  For the initial data corresponding to a real set of
astronomical observations, we find an estimated stability time of
$10^7$ years, which is not extremely smaller than the lifetime of the
Solar System ($\sim 5$~Gyr).
\end{abstract}


\end{frontmatter}

\section{Introduction}\label{sec:intro}
In this paper we revisit the problem of the stability of the Solar
System, at least considering (some of) the major planets, in the light
of both Kolmogorov and Nekhoroshev theories.  Some aspects are
also related to the theory of Lagrange and Laplace on the secular
motions of perihelia and nodes of the planetary orbits.

One of our main aims is to point out the major dynamical and
computational difficulties that arise in the application of
Kolmogorov's theorem.  In view of this, we  attempt to
apply the Nekhoroshev theory by trying essentially an extension of
Lagrange theory.  Although the final results appear to be
interesting, our conclusion will be that further and more refined
investigations are needed.  We consider indeed the present paper as the
beginning of a more comprehensive study of systems with more than two
planets in the framework of perturbation methods related to the
theories above.

In 1954 Kolmogorov announced his celebrated theorem on the persistence
under small perturbation of quasi periodic motions on invariant tori
of an integrable Hamiltonian systems (see~\cite{Kolmogorov-1954}).  The
relevance of that result for the problem of stability of the Solar
System was pointed out by Kolmogorov himself, and later emphasized in
the subsequent papers of Moser (see~\cite{Moser-1962}) and
Arnold (see~\cite{Arnold-1963}).  The three papers mentioned above marked
the beginning of the so called KAM theory.

However the actual applicability of Kolmogorov's theorem to the
planetary system encounters two major difficulties, namely: (i)~the
degeneracy of the Keplerian motion, and (ii)~the extremely restrictive
assumptions on the smallness of the perturbation.

The former difficulty is related  to the
elliptic form of the Keplerian orbits.  Indeed a system including a
central body (a star) and $n>1$ planets, after elimination of the
known first integrals, has $3n-2$ degrees of freedom, while only $n$
actions appear in the Keplerian part of the Hamiltonian.  The way out
proposed by Arnold, and inspired by the approach of Lagrange and
Laplace, was to introduce in the proof two separate time scales for
the orbital motion and for the secular evolution of the perihelia and
of the nodes (see~\cite{Arnold-1963.1} and its recent extension
in~\cite{Chi-Pinz-2009}).  Such an approach has been successfully
extended to the $n+1$-body planetary systems thanks to the work done
by Herman and F\'ejoz (see~\cite{Fejoz-2005}).

Attacking the second difficulty (i.e., the unrealistic requirements on
the smallness of masses, eccentricities and inclinations of the
planets) with purely analytical methods seems to be unrealistic.
However, some positive results could be attained using computer
algebra.  This means that we explicitly perform a few perturbation
steps, thus getting an approximation of the wanted invariant torus
which is good enough to allow us to apply analytical methods.  Such an
approach (also implementing interval arithmetic) allowed some authors
to rigorously prove the existence of KAM tori for some interesting
problems in Celestial Mechanics (see, e.g.,~\cite{Celletti-1994},
\cite{Cel-Chi-97}, \cite{Cel-Chi-2007}, \cite{Loc-Gio-2000}
and~\cite{Gab-Jor-Loc-2005}).  However, all these works consider
models having just two degrees of freedom.  This because increasing
the number of independent variables makes the explicit calculation of
perturbation steps a big challenge, due to the dramatic increase of
the number of coefficients to be calculated, so that a sufficiently
good initial approximation of an invariant torus is hardly obtained.
For what concerns problems with more than two degrees of freedom, in a
few cases only the availability of an algorithmic version of
Kolmogorov's theorem
(see~\cite{Ben-Gal-Gio-Str-84},~\cite{Giorgilli-1997.1}
and~\cite{Gio-Loc-1997.1}) allowed us to obtain a good approximation
of the invariant tori, although this approach is not yet sufficient
for a fully rigorous application of the theory.  For instance, the
constructed solution on a KAM torus has been successfully compared
with the real motion of the Sun--Jupiter--Saturn system, which can be
represented by a model with~$4$ degrees of freedom
(see~\cite{Loc-Gio-2007} for all details). Moreover our recent work
focuses on a first study of the long time stability in a neighborhood
of such KAM torus (see~\cite{Gio-Loc-San-2009}).

Besides the technical difficulty, the results of the numerical
explorations have raised some doubts concerning the
applicability of Kolmogorov theory to the major bodies of our
planetary system, namely the Sun and the so called Jovian planets,
i.e. Jupiter, Saturn, Uranus and Neptune, hereafter we will refer to
this model as the SJSUN problem. Indeed, the motion of such planetary
subsystem has been shown to be chaotic by Sussman and Wisdom
(see~\cite{Sus-Wis-1992}).  Murray and Holman provided such an
enlightening explanation of this phenomenon, that we think it is
helpful to briefly summarize some of their results as follows
(see~\cite{Mur-Hol-1999} for completeness). 

\begin{enumerate}
\renewcommand{\itemsep}{0pt}
\item[(a)] The chaoticity of the Jovian planets appears to be due
  to the {\it overlap of some resonances involving three or four
  bodies}.   An example is given by the resonances
  $$
  3n_1-5n_2-7n_3+\left[(3-j)g_1+6g_2+jg_3\right]\,,
  \qquad
  {\rm with}\ j=0,\,1,\,2,\,3\ ,
  $$ where $n_i$ stands for the mean motion frequency of the $i$-th
  planet, $g_i$ means the (secular) frequency of its perihelion
  argument and the indexes $1,\,2,\,3$ refer to Jupiter, Saturn and
  Uranus, respectively.  In fact, during the planetary motion each
  angle corresponding to the resonances above moves from libration to
  rotation and viceversa.  Many other resonances analogous to the
  previous ones are located in the vicinity of the real orbit of the
  SJSUN system, some of them involving also Neptune and the
  frequencies related to the longitudes of the nodes.
\item[(b)] The time needed by these resonances to eject Uranus from
  the Solar System is roughly evaluated to be about $10^{18}$ years.
\item[(c)] By moving the initial semi-major axis of Uranus in the
  range $19.18$--$19.35$ AU one observes some regions that look filled
  by quasi-periodic ordered motions and other regions that are weakly
  chaotic, i.e., with a Lyapunov time ranging between $2\times 10^5$
  and $10^8$ years.  All the main resonances acting in this region
  involve the linear combination $3n_1-5n_2-7n_3$ among the mean
  motion frequencies of Jupiter, Saturn and Uranus.
\item[(d)] The result~(c) qualitatively persists also for the {\it
  planar} SJSUN system or when the influence of Neptune is neglected.
\item[(e)] Conversely, no chaotic motions are detected in the {\it
  planar} system including the Sun, Jupiter, Saturn and Uranus
  (hereafter, SJSU for shortness) for the same initial values of the
  semi-major axis of Uranus considered at point~(c).  This suggests
  that the resonances described at point~(a) affect observable
  regions only when combined with some effects induced by Neptune or
  by the mutual inclinations.
\end{enumerate}

\noindent
By the way, we note that the resonances involving the linear
combination $3n_1-5n_2-7n_3$ are clearly related to the approximate
ratio $5:2$ and $7:1$ between the orbital motion of Jupiter and Saturn
and of Jupiter and Uranus, respectively.  Similarly, the ratio $2:1$
between Uranus and Neptune appears also to be relevant (historically,
this helped Le~Verrier to predict the existence and the location of
Neptune).  The low order of the latter quasi-resonance may explain why the
influence of Neptune induces some chaotic behavior, as pointed out
in~(d) and~(e) above.  An overlapping of resonances
  involving the linear combination $n_1-3n_2+3n_3-3n_4$ among the mean
  motion frequencies of all the Jovian planets has been actually
  detected in a small region including their orbits; this has been
  made possible by a combined use of some refined numerical
  investigation methods (see~\cite{Guzzo-2005} and~\cite{Guzzo-2006}).
  Moreover, the coexistence of quasi-periodic and chaotic motions in a
  neighborhood of the real initial conditions (already pointed out in
  the remark~(c) above) has been recently shown to be extremely
  complicated (see~\cite{Hayes-2007} and~\cite{Hayes-2008}).

The weak chaos in the motion of the Jovian planets (see,
  e.g.,~\cite{Khol-Kuz-2007} for a review on this topic)  makes
somehow hopeless the task of describing their long-term evolution by
a quasi-periodic approximation, as it is provided by the KAM theory.
Therefore it appears to be more natural to look for exponential
stability as assured by Nekhoroshev theory
(see~\cite{Nekhoroshev-1977} and~\cite{Nekhoroshev-1979}).  Indeed the
theorem of Nekhoroshev applies to an open set of initial conditions,
and states that the stability time increases exponentially with the
inverse of the perturbation parameter.  Our aim is to investigate
whether the SJSU system may remain close to its current conditions for
a time that exceeds the lifetime of the system itself; e.g., in our
case the age of the Universe, which is estimated to be $\sim 1.4\times
10^{10}$ years, could be enough.  We stress that the rather long time
reported in~(b) concerning the possible dissolution of the SJSUN
system seems to support our hope.  The approach based on Nekhoroshev
theory has been applied during the last decades to the case of the
Trojan asteroids, producing realistic results (see,
e.g.,~\cite{Giorgilli-1997}, \cite{Sko-Dok-2001}, \cite{Eft-San-2005}
and~\cite{Lho-Eft-Dvo-2008}).  Concerning the SJSUN system, we expect
that a combination of both the KAM and the Nekhoroshev theory could
prove that the motion remains close to an invariant torus for very
long times (see~\cite{Mor-Gio-1995} and~\cite{Gio-Loc-San-2009}).

In the present paper, we restrict our attention to the SJSU {\it
planar} system, due to the huge computational difficulties one
encounters during the expansion of the Hamiltonian.  Indeed, a rather
long preliminary work is necessary in order to give the Hamiltonian a
convenient form for starting more standard perturbation methods
(see~\cite{Loc-Gio-2000}, \cite{Loc-Gio-2005}
and~\cite{Loc-Gio-2007}).  We devote sects.~\ref{sec:2D_plan_Ham}
and~\ref{sbs:Kolm-like-transf} to this part of the problem.

Furthermore, in the line of Lagrange theory, we focus only on the
secular part of the Hamiltonian, which is derived in
subsect.~\ref{sbs:red-sec-Ham}.  Let us emphasize that all along both
sects.~\ref{sec:2D_plan_Ham} and~\ref{sec:secular-model} we pay a
special attention to include all the relevant terms related to the
three-body mean motion quasi-resonance $3n_1-5n_2-7n_3$ in view of
the remarks reported at points~(a) and~(c) above).

The secular system turns out to have the form of a perturbed system of
harmonic oscillators.  Let us recall that the stability of its
equilibrium point could be investigated using the theorem of
Dirichlet.  In fact, we can use as Lyapunov function either the
secular Hamiltonian or another integral of motion that is related to
the total angular momentum. For planar systems, the secular
Hamiltonian is a first integral with a minimum corresponding to the
case where all orbits are circular.  This is enough to ensure that the
eccentricities remain bounded forever if their initial values are small
enough. Indeed, Lagrange and Laplace actually proved Dirichlet's
theorem for the special case of the secular part of the spatial
planetary system, by using the third component of the total angular
momentum as Lyapunov function.  However, a chaotic evolution inside the
stability domain is not excluded by their approach: the eccentricity
or the inclination of a planet could increase up to the maximum value
allowed. This is in contrast with the features of the motion of the
Jovian planets, which have been discussed above.  Thus, even when we
limit ourselves to the study of the secular planar case, we think that
it is useful to proceed by investigating the stability into the light
of Nekhoroshev theory, since it provides also the information that the
eccentricities of the planets essentially have a quasi-periodic
behavior, apart from the very small contribution induced by the high
order perturbing terms. Moreover, our approach can be extended to more
refined models including the dependency on the mean motion angles. The
study of the secular Hamiltonian by using the Nekhoroshev theory is
worked out in sect.~\ref{sec:stab-sec-model}.

Finally, sect.~\ref{sec:conclu} is devoted to the conclusions.

\section{Classical expansion of the planar
  planetary Hamiltonian}\label{sec:2D_plan_Ham}

Let us consider four point bodies $P_0,\,P_1,\,P_2,\,P_3$, with masses
$m_0,\,m_1,\,m_2,\,m_3$, mutually interacting according to Newton's
gravitational law.  Hereafter the indexes $0,\,1,\,2,\,3$ will
correspond to Sun, Jupiter, Saturn and Uranus, respectively.

Let us now recall how the classical Poincar\'e variables can be
introduced so to perform a first expansion of the Hamiltonian around
circular orbits, i.e., having zero eccentricity.  We basically follow
the formalism introduced by Poincar\'e (see~\cite{Poincare-1892}
and~\cite{Poincare-1905}; for a modern exposition, see,
e.g.,~\cite{Laskar-1989b} and~\cite{Las-Rob-1995}).  We remove the
motion of the center of mass by using heliocentric coordinates
$\vet{r}_j=\build{P_0P_j}_{}^{\longrightarrow}\,$, with
$j=1,\,2,\,3\,$.  Denoting by $\tilde{\vet{r}}_j$ the momenta
conjugated to $\vet{r}_j$, the Hamiltonian of the system has $6$
degrees of freedom, and reads
\begin{equation}
F(\tilde{\vet{r}},\vet{r})=
T^{(0)}(\tilde{\vet{r}})+U^{(0)}(\vet{r})+
T^{(1)}(\tilde{\vet{r}})+U^{(1)}(\vet{r}) \ ,
\label{Ham-iniz}
\end{equation}
where
$$
\begin{array}{rclrcl}
T^{(0)}(\tilde{\vet{r}}) &= &\frac{1}{2}\build{\sum}_{j=1}^{3} 
\frac{m_0+m_j}{m_0m_j}\,\|\tilde{\vet{r}}_j\|^2\ ,
\qquad
&T^{(1)}(\tilde{\vet{r}}) &=
&\frac{1}{m_0}\Big(\tilde{\vet{r}}_1\cdot\tilde{\vet{r}}_2+
\tilde{\vet{r}}_1\cdot\tilde{\vet{r}}_3+
\tilde{\vet{r}}_2\cdot\tilde{\vet{r}}_3\Big)\ ,
\cr\cr
U^{(0)}(\vet{r}) &= &-\Gscr\build{\sum}_{j=1}^{3}
\frac{m_0\, m_j}{\|\vet{r}_j\|}\ ,
\qquad
&U^{(1)}(\vet{r})
&= &-\Gscr\left(\frac{m_1\, m_2}{\|\vet{r}_1-\vet{r}_2\|}
+\frac{m_1\, m_3}{\|\vet{r}_1-\vet{r}_3\|}
+\frac{m_2\, m_3}{\|\vet{r}_2-\vet{r}_3\|}\right)\ .
\cr
\end{array}
$$

The plane set of Poincar\'e's canonical variables is introduced as
\begin{equation}
\vcenter{\openup1\jot\halign{
 \hbox {\hfil $\displaystyle {#}$}
&\hbox {\hfil $\displaystyle {#}$\hfil}
&\hbox {$\displaystyle {#}$\hfil}
&\hbox to 6 ex{\hfil$\displaystyle {#}$\hfil}
&\hbox {\hfil $\displaystyle {#}$}
&\hbox {\hfil $\displaystyle {#}$\hfil}
&\hbox {$\displaystyle {#}$\hfil}\cr
\Lambda_j &=& \frac{m_0\, m_j}{m_0+m_j}\sqrt{\Gscr(m_0+m_j) a_j}\ ,
& &\lambda_j &=& M_j+\omega_j\ ,
\cr
\csi_j &=& \sqrt{2\Lambda_j}
\sqrt{1-\sqrt{1-e_j^2}}\,\cos\omega_j\ ,
& &\eta_j&=&-\sqrt{2\Lambda_j}
\sqrt{1-\sqrt{1-e_j^2}}\, \sin\omega_j\ ,
\cr
}}
\label{var-Poincare-piano}
\end{equation}
for $j=1\,,\,2\,,\,3\,$, where $a_j\,,\> e_j\,,\> M_j$ and $\omega_j$
are the semi-major axis, the eccentricity, the mean anomaly and the
perihelion longitude, respectively, of the $j$-th planet.  One
immediately sees that both $\csi_j$ and $\eta_j$ are of the same order
of magnitude as the eccentricity $e_j\,$.

Using Poincar\'e's variables~(\ref{var-Poincare-piano}), the
Hamiltonian $F$ can be rearranged so that one has
\begin{equation}
F(\vet{\Lambda},\vet{\lambda},\vet{\csi},\vet{\eta})=
F^{(0)}(\vet{\Lambda})+
\mu F^{(1)}(\vet{\Lambda},\vet{\lambda},\vet{\csi},\vet{\eta}) \ ,
\label{Ham-iniz-Poincare-var}
\end{equation}
where $F^{(0)}=T^{(0)}+U^{(0)}$, $\mu F^{(1)}=T^{(1)}+U^{(1)}$.  Here,
the small dimensionless parameter
$\mu=\max\{m_1\,/\,m_0\,$, $\,m_2\,/\,m_0\,$, $\,m_3\,/\,m_0\,\}$ has been
introduced in order to highlight the different size of the terms
appearing in the Hamiltonian.  Let us remark that the time derivative
of each coordinate is $\Oscr(\mu)$ but in the case of the angles
$\vet{\lambda}\,$. Therefore, according to the common language in
Celestial Mechanics, in the following we will refer to $\vet{\lambda}$
and to their conjugate actions $\vet{\Lambda}$ as the {\em fast
variables}, while $(\vet{\csi},\vet{\eta})$ will be called {\em
secular variables}.

We proceed now by expanding the
Hamiltonian~(\ref{Ham-iniz-Poincare-var}) in order to construct the
first basic approximation of Kolmogorov's normal form.  We pick a
value $\vet{\Lambda}^*$ for the fast actions and perform a translation
$\Tscr_{\vet{\Lambda}^*}$ defined as
\begin{equation}
L_j=\Lambda_j-\Lambda_j^*\ ,
\qquad{\rm for}\ j=1\,,\, 2\,,\, 3\,.
\label{def-L}
\end{equation}
This is a canonical transformation that leaves the coordinates
$\vet{\lambda}\,$, $\vet{\csi}$ and $\vet{\eta}$ unchanged.  The
transformed Hamiltonian
$\Hscr^{(\Tscr)}=F\circ\Tscr_{\vet{\Lambda}^*}\,$ can be expanded in
power series of $\vet{L},\,\vet{\csi},\,\vet{\eta}$ around the origin.
Thus, forgetting an unessential constant we rearrange the Hamiltonian
of the system as
\begin{equation}
\Hscr^{(\Tscr)}(\vet{L},\vet{\lambda},\vet{\csi},\vet{\eta})=
\vet{n}^*\cdot\vet{L}+
\sum_{j_1=2}^{\infty}h_{j_1,0}^{({\rm Kep})}(\vet{L})+
\mu\sum_{j_1=0}^{\infty}\sum_{j_2=0}^{\infty}
h_{j_1,j_2}^{(\Tscr)}(\vet{L},\vet{\lambda},\vet{\csi},\vet{\eta}) \ ,
\label{Ham-trasl-fast}
\end{equation}
where the functions $h_{j_1,j_2}^{(\Tscr)}$ are homogeneous
polynomials of degree $j_1$ in the actions $\vet{L}$ and of degree
$j_2$ in the secular variables $(\vet{\csi},\vet{\eta})\,$. The
coefficients of such homogeneous polynomials do depend analytically
and periodically on the angles $\vet{\lambda}\,$.  The terms
$h_{j_1,0}^{({\rm Kep})}$ of the Keplerian part are homogeneous
polynomials of degree $j_1$ in the actions $\vet{L}\,$, the explicit
expression of which can be determined in a straightforward manner.  In
the latter equation the term which is both linear in the actions and
independent of all the other canonical variables (i.e.,
$\vet{n}^*\cdot\vet{L}$) has been separated in view of its relevance
in perturbation theory, as it will be discussed in the next section.
We also expand the coefficients of the power series
$h_{j_1,j_2}^{(T_F)}$ in Fourier series of the angles
$\vet{\lambda}\,$.  The expansion of the Hamiltonian is a traditional
procedure in Celestial Mechanics.  We work out these expansions for
the case of the planar SJSU system using a specially devised algebraic
manipulation.  The calculation is based on the approach described in
sect.~2.1 of~\cite{Loc-Gio-2000}, which in turn uses the scheme
sketched in sect.~3.3 of~\cite{Robutel-1995}.

\begin{table*}
\caption[]{Masses $m_j$ and initial conditions for Jupiter, Saturn and
  Uranus in our planar model.  We adopt the AU as unit of length, the
  year as time unit and set the gravitational constant
  $\Gscr=1\,$. With these units, the solar mass is equal to
  $(2\pi)^2$. The initial conditions are expressed by the usual
  heliocentric planar orbital elements: the semi-major axis $a_j\,$,
  the mean anomaly $M_j\,$, the eccentricity $e_j$ and the perihelion
  longitude $\omega_j\,$.  The data are taken by JPL at the Julian
  Date $2440400.5\,$.  }
\label{tab:parameters_2D_SJSU}
\begin{center}
  \begin{tabular}{|c|l|l|l|}
\hline
& Jupiter ($j=1$) & Saturn ($j=2$) & Uranus ($j=3$)
\\
\hline
$m_{j}^{\phantom{\displaystyle 1}}$
& $(2\pi)^2/1047.355$
& $(2\pi)^2/3498.5$
& $(2\pi)^2/22902.98$
\\
$a_j$
& $5.20463727204700266$ & $9.54108529142232165$ & $19.2231635458410572$
\\
$M_j$
& $3.04525729444853654$ & $5.32199311882584869$ & $0.19431922829271914$
\\
$e_j$
& $0.04785365972484999$ & $0.05460848595674678$ & $0.04858667407651962$
\\
$\omega_j$
& $0.24927354029554571$ & $1.61225062288036902$ & $2.99374344439246487$ 
\\
\hline
\end{tabular}
\end{center}
\end{table*}

The reduction to the planar case is performed as follows.  We
pick from Table~IV of~\cite{Standish-1998} the initial conditions of
the planets in terms of heliocentric positions and velocities at the
Julian Date $2440400.5\,$.  Next, we calculate the corresponding
orbital elements with respect to the invariant plane (that is
perpendicular to the total angular momentum).  Finally we include the
longitudes of the nodes $\Omega_j$ (which are meaningless in the
planar case) in the corresponding perihelion longitude $\omega_j$ and
we eliminate the inclinations by setting them equal to zero.  The
remaining initial values of the orbital elements are reported in
Table~\ref{tab:parameters_2D_SJSU}.

Having determined the initial conditions we come to determining the
average values $(a_1^*\,,\,a_2^*\,,\,a_3^*)$ of the semi-major axes
during the evolution.  To this end we perform a long-term numerical
integration of Newton's equations starting from the initial conditions
related to the data reported in Table~\ref{tab:parameters_2D_SJSU}.
After having computed $(a_1^*\,,\,a_2^*\,,\,a_3^*)\,$, we determine
the values $\vet{\Lambda}^*$ via the first equation
in~(\ref{var-Poincare-piano}).  This allows us to perform the
expansion~(\ref{Ham-trasl-fast}) of the Hamiltonian as a function of
the canonical coordinates
$(\vet{L},\vet{\lambda},\vet{\csi},\vet{\eta})$.  In our calculations
we truncate the expansion as follows.  (a)~The Keplerian part is
expanded up to the quadratic terms.  The terms $h_{j_1,j_2}^{(\Tscr)}$
include: (b1)~the linear terms in the actions $\vet{L}\,$, (b2)~all
terms up to degree~$18$ in the secular variables
$(\vet{\csi},\vet{\eta})\,$, (b3)~all terms up to the trigonometric
degree $16$ with respect to the angles $\vet{\lambda}\,$.  Our choice
of the limits will be fully motivated in the next section.  However,
let us anticipate here that we focus our attention on the features of
our final secular model: we aim to include there as much as possible
of the effects of the quasi-resonances involving the mean motion
frequencies of the three planets the impact of which on the dynamics
has been discussed in the introduction.  Thus, we push our expansion
in $(\vet{\csi},\vet{\eta})$ and $\vet{\lambda}$ as high as possible,
although a low order truncation in $\vet{L}$ is used (see,
e.g.,~\cite{Khol-Gre-Kuz-2001} for some standard evaluation criteria).

\section{The  secular model}\label{sec:secular-model}
We look now for a good description of the secular dynamics.  A
straightforward method would be to include in the unperturbed
Hamiltonian also the average of the perturbation over the fast angles.
However, it has been remarked by Robutel (see~\cite{Robutel-tesi})
that the frequencies of the quasi-periodic flow given by this secular
Hamiltonian (often called of order one in the masses) are quite
different from the true ones.  The reason lies in the effect of the
mean motion quasi-resonance $5:2\,$.  Therefore we look for an
approximation of the secular Hamiltonian up to order two in the masses
(see, e.g.,~\cite{Laskar-1988}, \cite{Laskar-1989c},
\cite{Robutel-tesi}, \cite{Loc-Gio-2000}, \cite{Kuz-Khol-2006}
and~\cite{Lib-Hen-2007}).  To this end we follow the approach
in~\cite{Loc-Gio-2007}, carrying out two ``Kolmogorov-like''
normalization steps in order to eliminate the main perturbation terms
depending on the fast angles $\vet{\lambda}\,$.  We concentrate our
attention on the quasi-resonant angles $2\lambda_1-5\lambda_2\,$,
$\lambda_1-7\lambda_3$ and $3\lambda_1-5\lambda_2-7\lambda_3$, which
are the most relevant ones for the dynamics.  Our aim is to replace
the orbit with zero eccentricity with a quasi-periodic one that takes
into account the effect of such quasi-resonances up to the second
order in the masses.  The procedure is a little cumbersome, and
requires two main steps that we describe in the next two subsections.

\subsection{Partial reduction of the perturbation}\label{sbs:Kolm-like-transf}
We emphasize that the Fourier expansion of the
Hamiltonian~(\ref{Ham-trasl-fast}) is generated just by terms due to
two-body interactions, and so harmonics including more than two fast
angles cannot appear.  Thus, at first order in the masses only
harmonics with the quasi-resonant angles $2\lambda_1-5\lambda_2$ and
$\lambda_1-7\lambda_3$ do occur.  Actually, harmonics with the
quasi-resonant angle $3\lambda_1-5\lambda_2-7\lambda_3$ are generated
by the first Kolmogorov-like transformation, but are of second order
in the masses, and should be removed by the second Kolmogorov-like
transformation described in the next section.

Let us go into details.  We denote by $\big\lceil f
\big\rceil_{\vet{\lambda};K_{F}}$ the Fourier expansion of a function
$f$ truncated so as to include only its harmonics
$\vet{k}\cdot\vet{\lambda}$ satisfying the restriction $0<|\vet{k}|\le
K_{F}\,$, being $|\vet{k}|=|k_1|+|k_2|+|k_3|\,$.  We also denote by
$\langle\cdot\rangle_{\vet{\lambda}}$ the average with respect to the
angles $\lambda_1\,$, $\lambda_2\,$, $\lambda_3\,$.  The canonical
transformations are using the Lie series algorithm (see,
e.g.,~\cite{Giorgilli-1995}).

We set $K_{F}=8$ and transform the Hamiltonian~(\ref{Ham-trasl-fast})
as $\hat \Hscr^{(\Oscr 2)}=\exp\lie{\mu\,\chi_{1}^{(\Oscr 2)}}\,
\Hscr^{(\Tscr)}$ with the generating function $\mu\,\chi_{1}^{(\Oscr
2)}(\vet{\lambda},\vet{\csi},\vet{\eta})$ determined by solving the
equation
\begin{equation}
\sum_{j=1}^{3}n^*_j
\frac{\partial\,\chi_{1}^{(\Oscr 2)}}{\partial \lambda_j}
+\sum_{j_2=0}^{6}\left\lceil h_{0,j_2}^{(\Tscr)}
\right\rceil_{\vet{\lambda};8}
(\vet{\lambda},\vet{\csi},\vet{\eta})=0\ .
\label{eqperchi1Oscr2}
\end{equation}
Notice that, by definition, $\big\langle\big\lceil f
\big\rceil_{\vet{\lambda};K_{F}}\big\rangle_{\vet{\lambda}}=0\,$,
which assures that equation~(\ref{eqperchi1Oscr2}) can be solved
provided the frequencies $(n_1^*\,,\,n_{2}^*\,,\,n_{3}^*)$ are not
resonant up to order $8\,$, as it actually occurs in our planar model
of the SJSU system.

The Hamiltonian $\hat \Hscr^{(\Oscr 2)}$ has the same form of
$\Hscr^{(\Tscr)}$ in~(\ref{Ham-trasl-fast}), with the functions
$h_{j_1,j_2}^{(\Tscr)}$ replaced by new ones, that we denote by
$\hat{h}_{j_1,j_2}^{(\Oscr 2)}$, generated by the expanding the Lie
series $\exp\lie{\mu\,\chi_{1}^{(\Oscr 2)}}\,\Hscr^{(\Tscr)}$ and by
gathering all the terms having the same degree both in the fast
actions and in the secular variables.

Now we perform a second canonical transformation $\Hscr^{(\Oscr
2)}=\exp\lie{\mu\,\chi_{2}^{(\Oscr 2)}}\,\hat \Hscr^{(\Oscr 2)}\,$,
where the generating function $\mu\,\chi_{2}^{(\Oscr
2)}(\vet{L},\vet{\lambda},\vet{\csi},\vet{\eta})$ (which is linear
with respect to $\vet{L}$) is determined by solving the equation
\begin{equation}
\sum_{j=1}^{3}n^*_j
\frac{\partial\,\chi_{2}^{(\Oscr 2)}}{\partial \lambda_j}
+\sum_{j_2=0}^{6}\left\lceil \hat{h}_{1,j_2}^{(\Oscr 2)}
\right\rceil_{\vet{\lambda};8}
(\vet{L},\vet{\lambda},\vet{\csi},\vet{\eta})=0\ .
\label{eqperchi2Oscr2}
\end{equation}
Again, the Hamiltonian $\Hscr^{(\Oscr 2)}$ can be written in a form
similar to~(\ref{Ham-trasl-fast}), namely
\begin{equation}
\Hscr^{(\Oscr 2)}(\vet{L},\vet{\lambda},\vet{\csi},\vet{\eta})=
\vet{n}^*\cdot\vet{L}+
\sum_{j_1=2}^{\infty} h_{j_1,0}^{({\rm Kep})}(\vet{L})+
\mu\sum_{j_1=0}^{\infty}\sum_{j_2=0}^{\infty}
h_{j_1,j_2}^{(\Oscr 2)}
(\vet{L},\vet{\lambda},\vet{\csi},\vet{\eta};\mu)\ .
\label{Ham-Omu^2}
\end{equation}
where the new functions $h_{j_1,j_2}^{(\Oscr 2)}$ are calculated as
previously explained for $\hat{h}_{j_1,j_2}^{(\Oscr 2)}\,$.  Moreover,
they still have the same dependence on their arguments as
$h_{j_1,j_2}^{(\Tscr)}$ in~(\ref{Ham-trasl-fast}).

If terms of second order in $\mu$ are neglected, then the Hamiltonian
$\Hscr^{(\Oscr 2)}$ possesses the secular $3$-dimensional invariant
torus $\vet{L}=\vet{0}$ and $\vet{\csi}=\vet{\eta}=\vet{0}$.  Thus,
in a small neighborhood of the origin of the fast actions
  and  for small eccentricities the solutions of the system
with Hamiltonian $\Hscr^{(\Oscr 2)}$ differ from those of its average
$\langle \Hscr^{(\Oscr 2)}\rangle_{\vet{\lambda}}$ by a quantity
$\Oscr(\mu^2)$.  In this sense the average of the
Hamiltonian~(\ref{Ham-Omu^2}) approximates the real dynamics of the
secular variables up to order two in the masses,  and due to
  the choice $K_{F}=8$  takes into account the quasi-resonances $5:2$
between Jupiter and Saturn and $7:1$ between Jupiter and Uranus.

In this part of the calculation we produce a truncated series which is
represented as a sum of monomials
$$
c_{\vet{j},\vet{k},\vet{r},\vet{s}}\,L_1^{j_1} L_2^{j_2} L_3^{j_3} 
 \,\xi_1^{r_1} \xi_2^{r_2} \xi_3^{r_3}
  \,\eta_1^{s_1} \eta_2^{s_2} \eta_3^{s_3}
   \,{\scriptstyle{{\displaystyle{\sin}}\atop{\displaystyle{\cos}}}}
     (k_1\lambda_1+k_2\lambda_2+k_3\lambda_3)\ .
$$ The truncated expansion of~$\Hscr^{(\Oscr 2)}$ contains
$94\,109\,751$ such monomials.  We truncate our expansion at degree 16
in the fast angles $\vet{\lambda}$ and at degree 18 in the slow
variables $\vet{\csi},\,\vet{\eta}$ (we shall justify this choice at
the end of the next section).
\subsection{Second approximation and reduction to the secular
  Hamiltonian}\label{sbs:red-sec-Ham} The huge number of coefficients
determined till now does not allow us to continue by keeping all of
them.  Therefore, in view that we plan to consider the secular system,
we perform a partial average by keeping only the main terms that
contain the quasi-resonant angle $3\lambda_1-5\lambda_2-7\lambda_3$.  More
precisely, we first consider the reduced Hamiltonian
\begin{equation}
\left\langle \Hscr^{(\Oscr 2)}\big|_{\vet{L}
 =\vet{0}}\,\right\rangle_{\vet{\lambda}}=
\mu\sum_{j_2=0}^{\infty} \big\langle
h_{0,j_2}^{(\Oscr 2)}(\vet{\csi},\vet{\eta};\mu)
 \big\rangle_{\vet{\lambda}}\ ,
\label{sec-Ham-Omu^2}
\end{equation}
namely we set $\vet{L}=\vet{0}\,$, which results in replacing the
orbit having zero eccentricity with a close invariant torus of the
unperturbed Hamiltonian, and average $\Hscr^{(\Oscr 2)}$ by removing
all the Fourier harmonics depending on the angles.  Next, we select in
$\Hscr^{(\Oscr 2)}$ the Fourier harmonics that contain the wanted
quasi-resonant angle $3\lambda_1-5\lambda_2-7\lambda_3$ and add them to the
Hamiltonian~(\ref{sec-Ham-Omu^2}).  Finally, we perform on the
resulting Hamiltonian the second Kolmogorov-like step.  With more
detail, this is the procedure, which is an adaptation of a scheme
already used in~\cite{Loc-Gio-2000}.

For $(j_1,j_2)\in\naturali^2$ we select the quasi-resonant terms
\begin{equation}
\vcenter{\openup1\jot\halign{
 \hbox {\hfil $\displaystyle {#}$}
&\hbox {\hfil $\displaystyle {#}$\hfil}
&\hbox {$\displaystyle {#}$\hfil}\cr
\mu^2 h_{j_1,j_2}^{({\rm q.r.})}
  (\vet{L},\vet{\lambda},\vet{\csi},\vet{\eta}) &=
&\mu\,\big\langle\, h_{j_1,j_2}^{(\Oscr 2)}
\,\exp\big[-{\rm i}(3\lambda_1-5\lambda_2-7\lambda_3)\big]
\,\big\rangle_{\vet{\lambda}}
\,\exp\big[{\rm i}(3\lambda_1-5\lambda_2-7\lambda_3)\big]\,+
\cr
& &\mu\,\big\langle\, h_{j_1,j_2}^{(\Oscr 2)}
\,\exp\big[{\rm i}(3\lambda_1-5\lambda_2-7\lambda_3)\big]
\,\big\rangle_{\vet{\lambda}}
\,\exp\big[-{\rm i}(3\lambda_1-5\lambda_2-7\lambda_3)\big]
\ .
\cr
}}
\label{def-filter-terms-Omu^2}
\end{equation}
Actually, this means that in our expression we just remove all
monomials but the ones containing the wanted quasi-resonant angle.
Using the selected terms we determine a generating function
$\mu^2\chi_{1}^{({\rm q.r.})}(\vet{\lambda},\vet{\csi},\vet{\eta})$ by
solving the equation
\begin{equation}
\sum_{j=1}^{3}n^*_j
\frac{\partial\,\chi_{1}^{({\rm q.r.})}}{\partial \lambda_j}
+\sum_{j_2=0}^{9} h_{0,j_2}^{({\rm q.r.})}(\vet{\lambda},\vet{\csi},\vet{\eta})=0\ .
\label{eqperchi1filt}
\end{equation}
Here we make the calculation faster by keeping only terms up to degree
$9$ in~$(\vet{\csi},\vet{\eta})\,$, this allows us to keep the more
relevant quasi-resonant contributions.  Then, still following the
procedure outlined in~\cite{Loc-Gio-2000}, we calculate only the
interesting part of the transformed Hamiltonian
$\exp\lie{\mu^2\,\chi_{2}^{({\rm q.r.})}}\,
\exp\lie{\mu^2\,\chi_{1}^{({\rm q.r.})}}\,\Hscr^{(\Oscr 2)}\,$, namely
we keep in the transformation only the part which is independent of
all the fast variables $(\vet{L},\vet{\lambda})\,$.  This produces the
secular Hamiltonian $\Hscr^{({\rm sec})}$, which satisfies the formal
equation $\big\langle\exp\lie{\mu^2\,\chi_{2}^{({\rm q.r.})}}\,
\exp\lie{\mu^2\,\chi_{1}^{({\rm q.r.})}}\,\Hscr^{(\Oscr
  2)}\big\rangle_{\vet{\lambda}}= \Hscr^{({\rm
    sec})}+\Oscr(\|\vet{L}\|)+o(\mu^4)\,$, where
\begin{equation}
\vcenter{\openup1\jot\halign{
 \hbox {\hfil $\displaystyle {#}$}
&\hbox {\hfil $\displaystyle {#}$\hfil}
&\hbox {$\displaystyle {#}$\hfil}\cr
\Hscr^{({\rm sec})}(\vet{\csi},\vet{\eta})
&= &\mu\sum_{j_2=0}^{\infty} \big\langle
h_{0,j_2}^{(\Oscr 2)}\big\rangle_{\vet{\lambda}}+
\mu^4\Bigg\langle\,\frac{1}{2}
\fastpoisson{\chi_{1}^{({\rm q.r.})}}
{\Lscr_{\mu^2\,\chi_{1}^{({\rm q.r.})}}h_{2,0}^{({\rm Kep})}}
+\cr
&
&\ \fastpoisson{\chi_{1}^{({\rm q.r.})}}{\sum_{j_2=0}^{\infty} h_{1,j_2}^{({\rm q.r.})}}
+\frac{1}{2}\secpoisson{\chi_{1}^{({\rm q.r.})}}
{\sum_{j_2=0}^{\infty} h_{0,j_2}^{({\rm q.r.})}}
\,\Bigg\rangle_{\vet{\lambda}}\ .
\cr
}}
\label{def-sec-Ham}
\end{equation}
Here, we denoted by $\fastpoisson{\cdot}{\cdot}$ and
$\secpoisson{\cdot}{\cdot}$ the terms of the Poisson bracket involving
only the derivatives with respect the variables
$(\vet{L},\vet{\lambda})$ and $(\vet{\csi},\vet{\eta})$, respectively.

The Hamiltonian so constructed is the secular one, describing the slow
motion of eccentricities and perihelia.  In view of D'Alembert rules
$\Hscr^{({\rm sec})}$ contains only terms of even degree and so the
lowest order significant term has degree 2 (see,
e.g.,~\cite{Poincare-1905} and
also~\cite{Kholshevnikov-1997}--\cite{Kholshevnikov-2001} for a modern
approach, suitable for evaluating the non-zero coefficients).  We have
determined the power series expansion of the Hamiltonian up to degree
18 in the slow variables.  In order to allow a comparison with other
expansions, we reported our results up to degree~4
in~$(\vet{\csi},\vet{\eta})$ in appendix~\ref{app:exp_sec_mod}.

We close this section with a few remarks which justify our choice of
the truncation orders.  The limits on the expansions in the fast
actions $\vet{L}$ have been illustrated at points~(a) and~(b1) at the
end of section~\ref{sec:2D_plan_Ham}, and they are the smallest ones
that are required in order to make the Kolmogorov-like normalization
procedure significant.  Since we want to keep the quasi-resonant
angles $2\lambda_1-5\lambda_2\,$, $\lambda_1-7\lambda_3$ and
$3\lambda_1-5\lambda_5-7\lambda_3\,$, we set the truncation order for
Fourier series to $16$, which is enough.  The choice to truncate the
expansion at degree $18$ in the secular variables
$(\vet{\csi},\vet{\eta})$ is somehow subtler.  In view of D'Alembert
rules the harmonics $2\lambda_1-5\lambda_2$ and $\lambda_1-7\lambda_3$
have coefficients of degree at least 3 and 6, respectively, in the
secular variables.  Furthermore, the quasi-resonant angle
$3\lambda_1-5\lambda_5-7\lambda_3$ does not appear initially in the
Hamiltonian, but is generated by Poisson bracket between the harmonics
$2\lambda_1-5\lambda_2$ and $\lambda_1-7\lambda_3$, which produces
monomials of degree $9$ in $(\vet{\csi},\vet{\eta})$.  Therefore, we
decided to calculate the generating functions $\chi_{1}^{(\Oscr 2)}$
and $\chi_{2}^{(\Oscr 2)}$ up to degree $9$ (recall
equations~(\ref{eqperchi1Oscr2}) and~(\ref{eqperchi2Oscr2})).
Finally, in the second Kolmogorov-like step we want to keep the
secular terms generated by the harmonic
$3\lambda_1-5\lambda_5-7\lambda_3$, which are produced by Poisson
bracket between monomials containing precisely this harmonic, and then
the result has maximum degree $18$ in $(\vet{\csi},\vet{\eta})$.  This
explains the final truncation order for the slow variables.

\section{Stability of the secular Hamiltonian model}\label{sec:stab-sec-model}
The lowest order approximation of the secular Hamiltonian
$\Hscr^{({\rm sec})}$, namely its quadratic term, is essentially the
one considered in the theory first developed by Lagrange
(see~\cite{Lagrange-1776}) and furtherly improved by Laplace
(see~\cite{Laplace-1772}, \cite{Laplace-1784} and~\cite{Laplace-1785})
and by Lagrange himself (see~\cite{Lagrange-1781},
\cite{Lagrange-1782}).  In modern language, we say that the origin of
the reduced phase space (i.e.,
$(\vet{\csi},\vet{\eta})=(\vet{0},\vet{0})\,$) is an elliptic
equilibrium point (for a review using a modern formalism, see sect.~3
of~\cite{Bia-Chi-Val-2006}, where a planar model of our Solar System
is considered).

It is well known that (under mild assumptions on the quadratic part of
the Hamiltonian which are satisfied in our case) one can find a linear
canonical transformation
$(\vet{\csi},\vet{\eta})=\Dscr(\vet{x},\vet{y})$ which diagonalizes
the quadratic part of the Hamiltonian, so that we may  write
  $\Hscr^{({\rm sec})}$ in the new coordinates as
\begin{equation}
H^{(0)}(\vet{x},\vet{y})=\sum_{j=0}^{3}\frac{\nu_j}{2}\left(x_j^2+y_j^2\right)+
H_{2}^{(0)}(\vet{x},\vet{y})+H_{4}^{(0)}(\vet{x},\vet{y})+
H_{6}^{(0)}(\vet{x},\vet{y})+\ldots\ ,
\label{Ham^0}
\end{equation}
where $\nu_j$ are the secular frequencies in the small oscillations
limit and $H_{2s}^{(0)}$ is a homogeneous polynomial of degree $2s+2$
in $(\vet{x},\vet{y})\,$.  The calculated values of
$(\nu_1,\nu_2,\nu_3)$ in our case are reported in
Table~\ref{tab:freq_condiniz_sec_2D_SJSU}.

\begin{table*}
\caption[]{Angular velocities $\vet{\nu}$ and initial conditions
  $(\vet{x}(0),\vet{y}(0))$ for our planar secular model about the
  motions of Jupiter, Saturn and Uranus. The frequency vector
  $\vet{\nu}$ refer to the harmonic oscillators approximation of
  the Hamiltonian $H^{(0)}$ (written in~(\ref{Ham^0})) and its values
  are given in $rad/year\,$.}
\label{tab:freq_condiniz_sec_2D_SJSU}
\begin{center}
\begin{tabular}{|c|l|l|l|}
\hline
& \hfil$j=1$ &\hfil$j=2$ & \hfil$j=3$
\\
\hline
$\nu_{j}^{\phantom{\displaystyle 1}}$
& $-1.1212724892\,\times 10^{-4}$
& $-1.9688444678\,\times 10^{-5}$
& $-1.1134564418\,\times 10^{-5}$
\\
$x_j(0)$
& $\phantom{-}1.5407573458\,\times 10^{-2}$
& $-3.0574059274\,\times 10^{-2}$
& $\phantom{-}1.1186486403\,\times 10^{-2}$
\\
$y_j(0)$
& $-2.5320810665\,\times 10^{-2}$
& $-5.2728862107\,\times 10^{-3}$
& $\phantom{-}6.0669645406\,\times 10^{-3}$
\\
\hline
\end{tabular}
\end{center}
\end{table*}

Thus, we are led to study the stability of the equilibrium for the
Hamiltonian~(\ref{Ham^0}).  As remarked in the introduction, perpetual
stability in a neighborhood of the equilibrium is ensured by applying
Dirichlet's theorem. We can do it in two ways. Since all frequencies
have the same sign (that is negative in our case), we can use
$H^{(0)}$ as Lyapunov function; actually, a very rough evaluation of
the size of the stability neighborhood gives us a value of the radius
that is about $0.6$ times the distance (from the origin) of the actual
initial data of the planets.  Such an estimate should certainly be
improved by a more accurate calculation, i.e., by determining the
stationary points of a function in 6 variables.  A second alternative
approach focuses on the constant of motion related to the total
angular momentum. In fact, $A=\sum_{j=1}^{3}(\csi_j^2+\eta_j^2)/2$ is
a first integral for the initial Hamiltonian
$F(\vet{\Lambda},\vet{\lambda},\vet{\csi},\vet{\eta})$ described
in~(\ref{Ham-iniz-Poincare-var}). Thus, one can use as Lyapunov
function the average of the angular momentum $A\,$, when it is
expressed as a function of the new secular canonical coordinates
$(\vet{x},\vet{y})$.  However, we emphasize that the Dirichlet theory
does not apply to a complete planetary system, that is not averaged
with respect to the mean motion angles.  Thus, as said in the
introduction, we think it is more interesting to investigate the
stability of the equilibrium in the light of Nekhoroshev theory, in
view of its possible extension to more refined models.

\subsection{Birkhoff's normal form}\label{sbs:Birkhoff}
Following a quite standard procedure we proceed to construct the
Birkhoff's normal form for the Hamiltonian~(\ref{Ham^0})
(see~\cite{Birkhoff-27}; for an application of Nekhoroshev theory
see, e.g.,~\cite{Giorgilli-1988}).  This is a well known matter, thus
we limit our exposition to a short sketch adapted to the present
context.

The aim is to give the Hamiltonian the normal form at order $r$
\begin{equation}
H^{(r)}(\vet{x},\vet{y})=Z_{0}(\vet{\Phi})+\ldots+Z_{r}(\vet{\Phi})+
\Fscr_{r+1}^{(r)}(\vet{x},\vet{y})+\Fscr_{r+2}^{(r)}(\vet{x},\vet{y})+\ldots\ ,
\label{Ham^r}
\end{equation}
where
\begin{equation}
\Phi_j=\frac{1}{2}\left(x_j^2+y_j^2\right)
\qquad{\rm for}\  j=1,\,2,\,3
\label{def-Phi}
\end{equation}
are the actions of the system, and $Z_{s}$ for $s=0,\,\ldots\,,\,r\,$
is a homogeneous polynomial of degree $s/2+1$ in $\vet{\Phi}$ and in
particular it is zero for odd $s$.  The un-normalized reminder terms
$\Fscr_{s}^{(r)}$, where $s>r\,$, are homogeneous polynomials of degree
$s+2$ in $(\vet{x},\vet{y})\,$.

We proceed by induction.  Assume that the Hamiltonian is in normal
form up to a given order $r$, which is trivially true for $r=0\,$, and
determine a generating function $\chi^{(r+1)}$ and the normal form
term $Z_{r+1}\,$, by solving the equation
\begin{equation}
\adaptpoisson{\chi^{(r+1)}}{\vet{\nu}\cdot\vet{\Phi}}+
\Fscr_{r+1}^{(r)}(\vet{x},\vet{y})=
Z_{r+1}(\vet{\Phi})\ .
\label{eqperchir+1}
\end{equation}
Using the algorithm of Lie series transform, we can write the new
Hamiltonian as $H^{(r+1)}=\exp\lie{\chi^{(r+1)}}\,H^{(r)}$.  It is not
difficult to show that $H^{(r+1)}$ has a form analogous to that
written in~(\ref{Ham^r}) with new functions $\Fscr_{s}^{(r+1)}$ of
degree $s+2$ (where $s>r+1$) and the normal form part ending with
$Z_{r+1}\,$, which is equal to zero if $r$ is even (see,
e.g.,~\cite{Gio-Gal-1978}).  As usual when using the Lie series
methods, we denote by $(\vet{x},\vet{y})$ the new coordinates, so that
the normal form $H^{(r)}$ possesses the approximate first integrals
$\vet{\Phi}$ given by~(\ref{def-Phi}).  By the way, the algorithm can
be iterated up to the step $r$ provided that the non-resonance
condition
\begin{equation}
\vet{k}\cdot\vet{\nu}\neq 0
\qquad\forall\ 
\vet{k}\in\interi^3\ {\rm such\ that}\ 0<|\vet{k}|\le r+2
\label{cond-non-ris}
\end{equation}
is fulfilled.  

\subsection{Study of the stability time}\label{sbs:stab-times}
It is well known that Birkhoff's normal form at any finite
order $r$ is convergent in some neighborhood of the origin, but the
analyticity radius shrinks to zero when $r\to\infty\,$.  Thus, the best
one can do is to look for stability for a finite but long time.  We
use the algorithm reported in~\cite{GLS-2010}, that we describe here.

Let us pick three positive numbers $R_1,\,R_2,\,R_3$ and
consider a polydisk $\Delta_{\rho\vet{R}}$ with center at the origin
of $\reali^{6}$ defined as
$$
\Delta_{\rho\vet{R}}=\left\{(\vet{x},\vet{y})\in\reali^6:
\,x_j^2+y_j^2\leq\rho^2 R_j^2\,,\  j=1,\,2,\,3
\right\}\ ,
$$ $\rho>0$ being a parameter.  Let $\rho_0=\rho/2\,$, and let
$(\vet{x}_0,\vet{y}_0)\in\Delta_{\rho_0\vet{R}}$ be the initial point
of an orbit, so that one has $\Phi_j(0) = (x_j^2+y_j^2)/2\leq\rho_0^2
R_j^2/2$.  Therefore, there is $T(\rho_0)>0$ such that for $|t|\le
T(\rho_0)$ we have $\vet{\Phi}(t)\leq\rho^2 R_j^2/2\,$, and so also
$(\vet{x}(t),\vet{y}(t))\in\Delta_{\rho\vet{R}}$.  We call $T(\rho_0)$
the \emph{estimated stability time}, and our  aim is to give
a good estimate of it.

The key remark is that one has 
\begin{equation}
\dot\Phi_j=\adaptpoisson{\Phi_j}{H^{(r)}}=
\sum_{s=r+1}^{\infty}\adaptpoisson{\Phi_j}{\Fscr_{s}^{(r)}}
\simeq\adaptpoisson{\Phi_j}{\Fscr_{r+1}^{(r)}}
\qquad
{\rm for}\ j=1,\,2,\,3\ ,
\label{eq:derPhi}
\end{equation}
which holds true for an arbitrary normalization order $r$.  This means
that the time derivative of $\vet{\Phi}(t)$ is small, being
$\Oscr(\rho^{r+3})$, so that the time $T(\rho_0)$ may grow very large.
The basis of Nekhoroshev theory is that one can choose an optimal
value of $r$ as a function of $\rho_0$ letting it to get larger and
larger when $\rho_0\to 0$, so that $T(\rho_0)$ grows faster than any
power of $1/\rho_0\,$.  Here we give this argument an algorithmic
form, thus producing an explicit estimate of $T(\rho_0)\,$.

Let us write a homogeneous polynomial $f(\vet{x},\vet{y})$ of degree
$s$ as 
$$
f(\vet{x},\vet{y}) =
\sum_{|\vet{j}|+|\vet{k}| = s} f_{\vet{j},\vet{k}} \vet{x}^{\vet{j}} \vet{y}^{\vet{k}}\ ,
$$
where the multiindex notation $\vet{x}^{\vet{j}} \vet{y}^{\vet{k}} =
x_1^{j_1} x_2^{j_2} x_3^{j_3}y_1^{k_1} y_2^{k_2} y_3^{k_3}$ has been
used.  We define the quantity $|f|_{\vet{R}}$ as
\begin{equation}
|f|_{\vet{R}} = \sum_{|\vet{j}|+|\vet{k}| = s} |f_{\vet{j},\vet{k}}| 
 R_1^{j_1+k_1} R_2^{j_2+k_2} R_3^{j_3+k_3} 
  \Theta_{j_1,k_1}\Theta_{j_2,k_2}\Theta_{j_3,k_3}
\ ,\quad
\Theta_{j,k} = \sqrt{\frac{j^j k^k}{(j+k)^{j+k}}}\ .
\label{frm:20}
\end{equation}
We claim that for  $\rho\gt 0$ one has
\begin{equation}
\sup_{(\vet{x},\vet{y})\in\Delta_{\rho\vet{R}}} \bigl|f(\vet{x},\vet{y})\bigr| 
 \lt \rho^s |f|_{\vet{R}}\ .
\label{normabbona}
\end{equation}
The estimate is checked as follows.  In the plane $x_i,\,y_i$ consider
a disk with radius $R_i$.  Then inside the disk the inequality
$|x_i^{j_i} y_i^{k_i}| \le R_i^{j_i+k_i}\Theta_{j_i,k_i}$ holds true.
In fact, after having set $x_i=R_i\cos\theta\,,\>y_i=R_i\sin\theta\,$, one
can easily check that
$|\cos^{j_i}\theta\sin^{k_i}\theta|\leq\Theta_{j_i,k_i}\,$.  It is
then straightforward to verify that for a monomial $\vet{x}^{\vet{j}}
\vet{y}^{\vet{k}}$ of degree $s$ one has
$$
\sup_{(\vet{x},\vet{y})\in\Delta_{\rho\vet{R}}}
\bigl|\vet{x}^{\vet{j}} \vet{y}^{\vet{k}}\bigr| \leq \rho^s R_1^{j_1+k_1} R_2^{j_2+k_2}
R_3^{j_3+k_3} \Theta_{j_1,k_1} \Theta_{j_2,k_2} \Theta_{j_3,k_3}\ .
$$
The wanted inequality is just the sum of the contributions of all
monomials.

Using~(\ref{normabbona}) and~(\ref{eq:derPhi}) we can estimate
\begin{equation}
\sup_{(\vet{x},\vet{y})\in\Delta_{\rho\vet{R}}} \bigl|\dot\Phi_j(x,y)\bigr| 
 \lt C\rho^{r+3}\bigl| \{\Phi_j, \Fscr_{r+1}^{(r)}\}\bigr|_{\vet{R}}
\label{frm:21}
\end{equation}
for $j=1,\,2,\,3$ and with some $C\geq 1\,$.  In fact,
after having set $\rho$ smaller than the convergence radius 
  of the remainder series $\Fscr_{s}^{(r)}$ (where $s>r$), the
above inequality is true for some $C\,$.  In our calculation we set
$C=2\,$.

We come now to the calculation of the estimated stability time.  Since
$\Phi_j =\rho^2R_j^2/2\,$, we have $\dot \Phi_j = R_j^2\rho\dot\rho$
and, in view of inequality~(\ref{frm:21}), also
$$
\dot\rho \le \frac{B_{r,j}}{R_j^2} \rho^{r+2}\ ,\quad
 B_{r,j} = C \bigl| \{\Phi_j, \Fscr_{r+1}^{(r)}\}\bigr|_{\vet{R}}\ .
$$ 
Thus a majorant of the function $\rho(t)$ is given by the solution
of the equation $\dot\rho = B_{r,j} \rho^{r+2}/ R_j^2\,$.
Setting $\rho_0$ as the initial value we conclude
that $\rho(t)\leq 2\rho_0$ for all $|t| \leq\tau(\rho_0,r)$, where
\begin{equation}
\tau(\rho_0,r) = \min_j
 \frac{R_j^2}{B_{r,j}} \int_{\rho_0}^{2\rho_0} \frac{d\sigma}{\sigma^{r+2}} = 
\min_j
 \left(1-\frac{1}{2^{r+1}}\right)
  \frac{R_j^2}{(r+1)B_{r,j}\,\rho_0^{r+1}}\ .
\label{frm:99}
\end{equation}
The latter estimate holds true for arbitrary normalization order $r$.
Therefore we select an optimal order  $r_{\rm opt}(\rho_0)$ by
looking for the maximum over $r$ of $\tau(\rho_0,r)$, thus getting
\begin{equation}
T(\rho_0) = \max_{r} \tau(\rho_0,2\rho_0,r)\ .
\label{frm:12}
\end{equation}
This is the best estimate of the stability time given by our algorithm.

\begin{figure}
\centering
\subfigure[Optimal normalization order]
{\includegraphics[width=90mm, angle=-90]{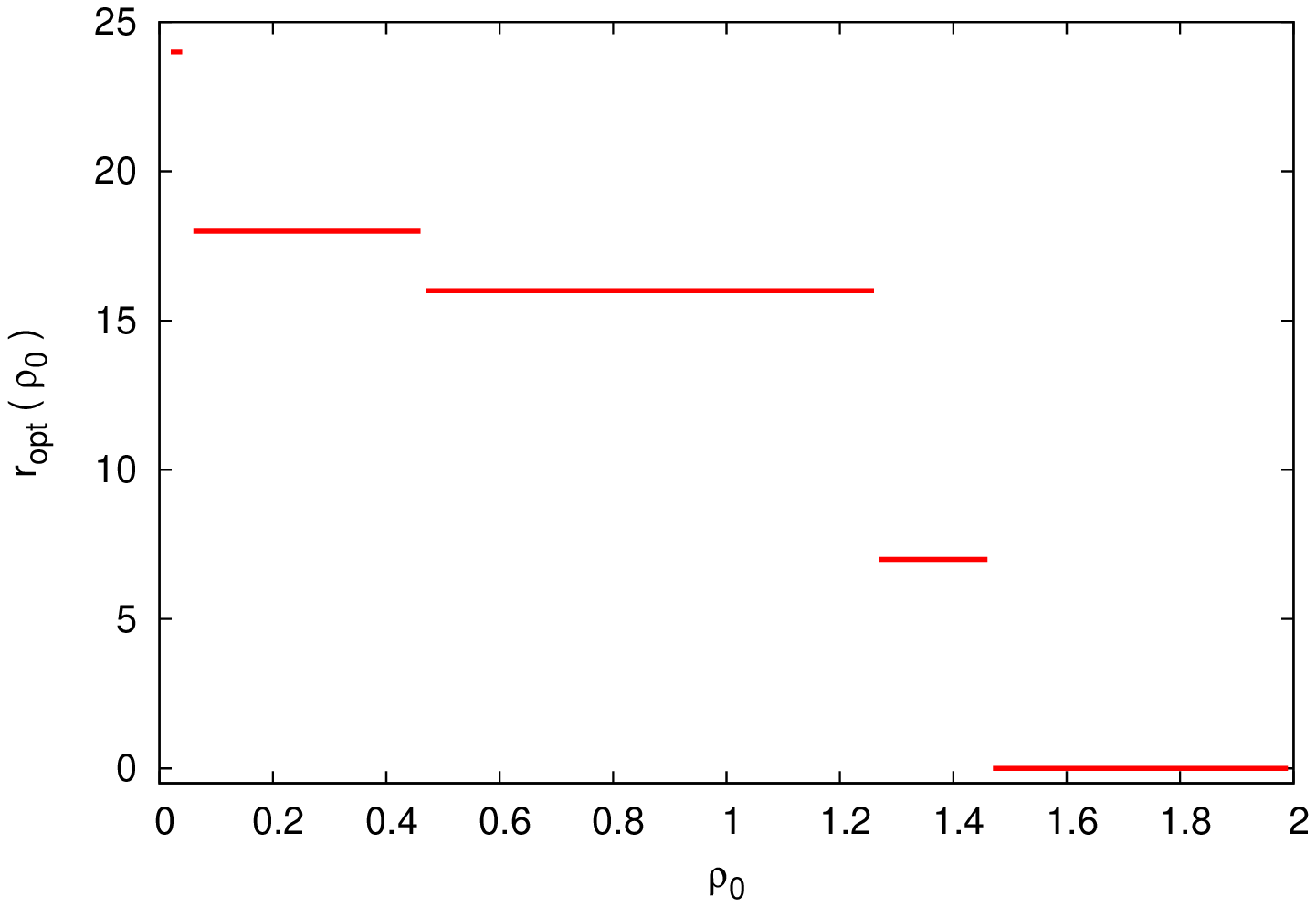}}
\subfigure[Estimated stability time]
{\includegraphics[width=90mm, angle=-90]{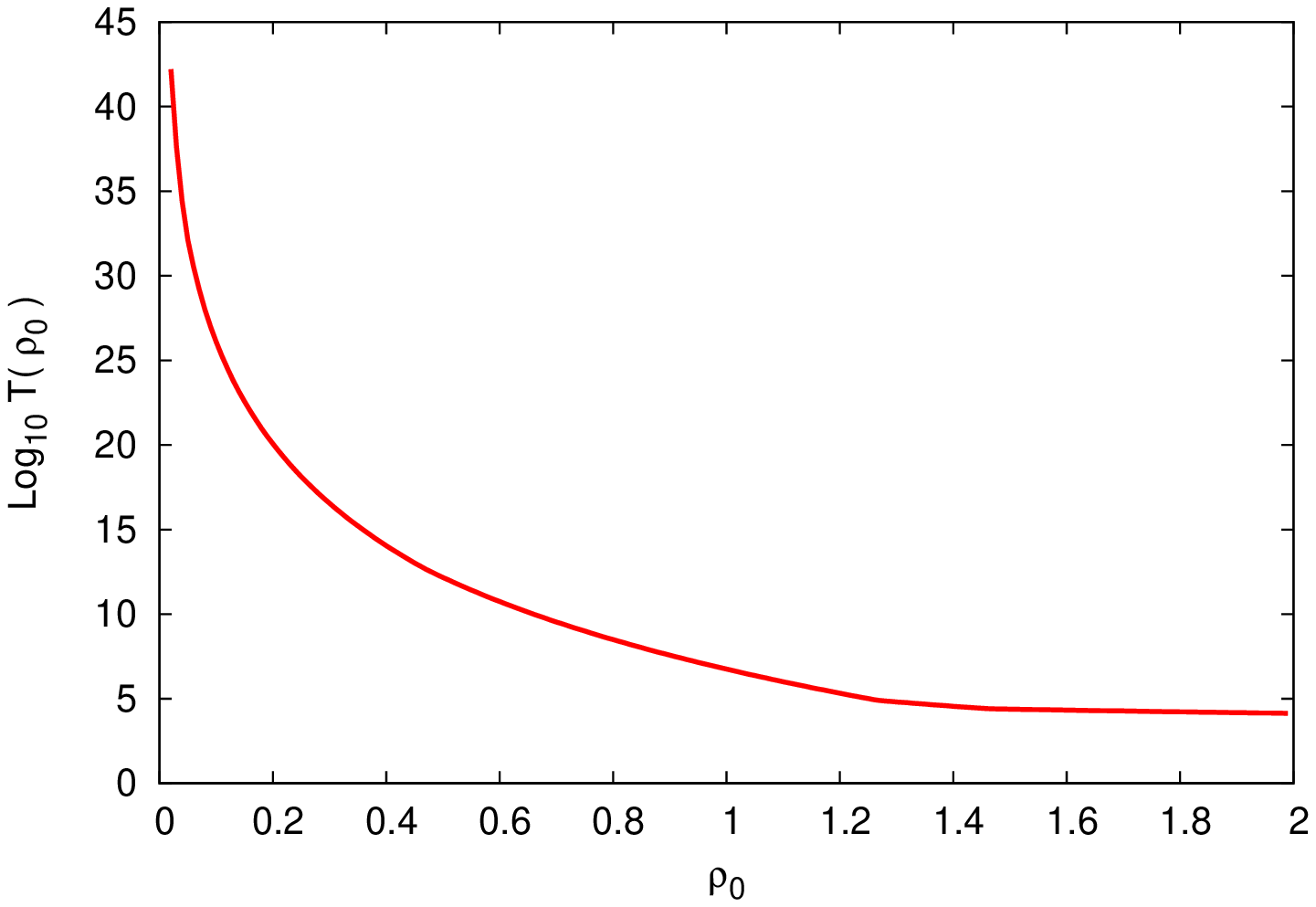}}
\caption{Optimal normalization order $r_{\rm opt}$ and estimated
  stability time $T(\rho_0)$ evaluated according to the algorithm of
  sect.~\ref{sbs:stab-times}.  The time unit is the year.  See text
  for more details.}
\label{fig:opt_norm_step}
\end{figure}

\subsection{Application to the SJSU system}
We apply the algorithm of the previous section to the secular
Hamiltonian $\Hscr^{({\rm sec})}$ by explicitly performing the
construction of Birkhoff's normal form up to order~$30$.  Meanwhile
also the first term of the remainder has been stored, so that the
estimate for $\dot{\vet{\Phi}}$ is provided.

The calculation of the estimated stability time is performed by
setting
\begin{equation}
R_1=2.5558203988\,\times 10^{-2} \ ,
\quad
R_2=3.0601862602\,\times 10^{-2} \ ,
\quad
R_3=1.1223294461\,\times 10^{-2} \ .
\label{def-Theta}
\end{equation}
These values have been calculated as $R_j = \sqrt{x_j^2(0)+y_j^2(0)}$
where $x_j(0),\,y_j(0)$ are the initial data reported in
table~\ref{tab:freq_condiniz_sec_2D_SJSU}, so that the initial point
is on the border of the polydisk $\Delta_{\rho\vet{R}}$ with $\rho=1\,$.

Finally we proceed to calculating the optimal normalization order
$r_{\rm opt}(\rho_0)$  and the estimated stability time
$T(\rho_0)$ as functions of $\rho_0$ in an interval such that the
optimal normalization order produced by our algorithm is less than
$30$.  The results are reported in fig.~\ref{fig:opt_norm_step}.  
The 
fast increase of the time when $\rho$ decreases is evident from the
graph.  We also remark that for $\rho_0=1$, which corresponds to the
initial data for the planets, the normalization order is already
$r_{\rm opt} =16$.  This shows that the mechanism of long time
stability is already active.  The estimated time with our algorithm is
about $10^7$ years for $\rho_0=1$.  This seems to be quite short both
with respect to the age of the Solar System (which is
estimated to be $\sim 5\times 10^{9}$ years) and with respect to the
numerical indications ($10^{18}$ years).  We shall comment on this
point in the next section.

\section{Conclusions and outlooks}\label{sec:conclu}
In the framework provided by the Nekhoroshev's theorem, the present
paper describes the first attempt to study the stability of a
realistic model with more than two planets of our Solar System.  As
remarked at the end of the previous section we are not yet able to
prove the stability for a time comparable to the age of our planetary
system, even restricting ourselves to consider just the secular part
of a planar approximation including the Sun, Jupiter, Saturn and
Uranus.  Nevertheless, we think that our result is meaningful in that
it indicates that the phenomenon of exponential stability in
Nekhoroshev sense may play an effective role for the Solar System, at
least for the biggest planets.  On the other hand, we stress that our
result is not dramatically far from the goal of proving stability for
the age of the Solar System: such a time is reached for $\rho_0\sim
0.7\,$.  By the way, it may be worth to note that a similar result,
with the same value of the radius, has been found in~\cite{GLS-2010}
where the spatial problem for the Sun--Jupiter--Saturn system is
considered.  Such a value of $\rho_0$ appears to be not so small,
especially if one recalls the rough estimates based on the first
purely analytical proofs of the KAM theorem: in order to apply them to
some model of our planetary system, the Jupiter mass should be smaller
than that of a proton.  Improvements are surely possible, and the
relatively short history of the applications of the Nekhoroshev type
estimates to Celestial Mechanics has shown definitely more remarkable
improvements than the one required here (e.g.,
compare~\cite{Giorgilli-1989} with~\cite{Giorgilli-1997}).

Some drawbacks are immediately evident.  The most relevant one is that
the estimate in~(\ref{frm:99}) actually assumes that the perturbation
constantly forces the worst possible evolution.  This is clearly
pessimistic, and justifies the striking difference with respect to the
indication given by the numerical integrations.  On the other hand,
general perturbation method are essentially based on estimates that
are often very crude.  The explicit calculation of normal forms and
related quantities allows us to significantly improve our results, but
the price is either a bigger and bigger computer power or more and
more refined methods. 

The natural question is whether there is a way to improve the present
result.  Our approach suggests that a better approximation of
the true orbit could help a lot.  This can be obtained, e.g., by first
establishing the existence of a KAM torus close to the initial
conditions of the planets, and then proving the stability in
Nekhoroshev sense in a neighborhood of the torus that contains the
initial data.  Such an approach has been attempted
in~\cite{Gio-Loc-San-2009} for the Sun--Jupiter--Saturn case
considering the full system, i.e., avoiding the approximation of the
secular model.  In that case the number of coefficients to be
handled is so huge that the calculation can actually be performed only
by introducing strong truncations on the expansions; this might
artificially improve the results.  Thus, some new idea is necessary,
and this will be work for the future.

\subsection*{Acknowledgments}
The authors have been supported by the research program ``Dynamical
Systems and applications'', PRIN 2007B3RBEY, financed by MIUR.
M.S. has been partially supported also by the research program ``Studi
di Esplorazione del Sistema Solare'', financed by ASI.

\appendix
\section{Expansion of the secular Hamiltonian of the planar SJSU
  system up to order 2 in the masses and 4 in
  eccentricities}\label{app:exp_sec_mod}

Our secular model is represented by the Hamiltonian $\Hscr^{({\rm
    sec})}$, which is defined in~(\ref{def-sec-Ham}).  Here, we limit
ourselves to report the expansion of $\Hscr^{({\rm sec})}$ up to
degree~$4$ in $(\vet{\csi},\vet{\eta})\,$. Therefore, as a consequence
of the D'Alembert rules, the terms related to the quasi-resonance
$3\lambda_1-5\lambda_2-7\lambda_3$ do not give any contribution to the
coefficients listed below. Thus, the following expansion of the rhs
of~(\ref{def-sec-Ham}) actually takes into account just
$\mu\big\langle h_{0,2}^{(\Oscr 2)}\big\rangle_{\vet{\lambda}}+
\mu\big\langle h_{0,4}^{(\Oscr 2)}\big\rangle_{\vet{\lambda}}$ (recall
that $\Hscr^{({\rm sec})}$ contains just terms of even degree in its
variables $(\vet{\csi},\vet{\eta})\,$). The calculation of the
functions $h_{0,2}^{(\Oscr 2)}$ and $h_{0,4}^{(\Oscr 2)}$ is performed
how it has been explained in subsect.~\ref{sbs:Kolm-like-transf}.
\begin{align*}
&\Hscr^{({\rm sec})}(\vet{\csi},\vet{\eta})=
\cr
&  -2.0438249530856989\times 10^{-05}\,\xi^{2}_{1}\,
&  +3.9042681895470743\times 10^{-05}&\,\xi^{1}_{1}\,\xi^{1}_{2}\,\cr
&  +4.5005164146422330\times 10^{-07}\,\xi^{1}_{1}\,\xi^{1}_{3}\,
&  -4.5352294644578622\times 10^{-05}&\,\xi^{2}_{2}\,\cr
&  +1.9490388069796070\times 10^{-06}\,\xi^{1}_{2}\,\xi^{1}_{3}\,
&  -5.6845848333331483\times 10^{-06}&\,\xi^{2}_{3}\,\cr
&  -2.0438249530856989\times 10^{-05}\,\eta^{2}_{1}\,
&  +3.9042681895470675\times 10^{-05}&\,\eta^{1}_{1}\,\eta^{1}_{2}\,\cr
&  +4.5005164146422409\times 10^{-07}\,\eta^{1}_{1}\,\eta^{1}_{3}\,
&  -4.5352294644578622\times 10^{-05}&\,\eta^{2}_{2}\,\cr
&  +1.9490388069796070\times 10^{-06}\,\eta^{1}_{2}\,\eta^{1}_{3}\,
&  -5.6845848333331441\times 10^{-06}&\,\eta^{2}_{3}\,\cr
&  -1.0838003720922759\times 10^{-04}\,\xi^{4}_{1}\,
&  +1.2014175808584642\times 10^{-03}&\,\xi^{3}_{1}\,\xi^{1}_{2}\,\cr
&  +6.2045352476790196\times 10^{-07}\,\xi^{3}_{1}\,\xi^{1}_{3}\,
&  -4.5563232782076350\times 10^{-03}&\,\xi^{2}_{1}\,\xi^{2}_{2}\,\cr
&  +8.8406443127175810\times 10^{-07}\,\xi^{2}_{1}\,\xi^{1}_{2}\,\xi^{1}_{3}\,
&  -9.7678628300067324\times 10^{-06}&\,\xi^{2}_{1}\,\xi^{2}_{3}\,\cr
&  -2.1676479523871672\times 10^{-04}\,\xi^{2}_{1}\,\eta^{2}_{1}\,
&  +1.2014125316196400\times 10^{-03}&\,\xi^{2}_{1}\,\eta^{1}_{1}\,\eta^{1}_{2}\,\cr
&  +6.2157409102827665\times 10^{-07}\,\xi^{2}_{1}\,\eta^{1}_{1}\,\eta^{1}_{3}\,
&  -1.5832006427474584\times 10^{-03}&\,\xi^{2}_{1}\,\eta^{2}_{2}\,\cr
&  +3.0033462029049336\times 10^{-07}\,\xi^{2}_{1}\,\eta^{1}_{2}\,\eta^{1}_{3}\,
&  -7.4173186653205456\times 10^{-06}&\,\xi^{2}_{1}\,\eta^{2}_{3}\,\cr
&  +7.6046689202847869\times 10^{-03}\,\xi^{1}_{1}\,\xi^{3}_{2}\,
&  -2.4429460187142667\times 10^{-06}&\,\xi^{1}_{1}\,\xi^{2}_{2}\,\xi^{1}_{3}\,\cr
&  +3.9912387029285291\times 10^{-07}\,\xi^{1}_{1}\,\xi^{1}_{2}\,\xi^{2}_{3}\,
&  +1.2014125316196422\times 10^{-03}&\,\xi^{1}_{1}\,\xi^{1}_{2}\,\eta^{2}_{1}\,\cr
&  -5.9464179266765730\times 10^{-03}\,\xi^{1}_{1}\,\xi^{1}_{2}\,\eta^{1}_{1}\,\eta^{1}_{2}\,
&  +5.8365071555190281\times 10^{-07}&\,\xi^{1}_{1}\,\xi^{1}_{2}\,\eta^{1}_{1}\,\eta^{1}_{3}\,\cr
&  +7.6047082339419673\times 10^{-03}\,\xi^{1}_{1}\,\xi^{1}_{2}\,\eta^{2}_{2}\,
&  -1.6360484568891480\times 10^{-06}&\,\xi^{1}_{1}\,\xi^{1}_{2}\,\eta^{1}_{2}\,\eta^{1}_{3}\,\cr
&  +2.2482538047243290\times 10^{-07}\,\xi^{1}_{1}\,\xi^{1}_{2}\,\eta^{2}_{3}\,
&  +2.6233130055605185\times 10^{-05}&\,\xi^{1}_{1}\,\xi^{3}_{3}\,\cr
&  +6.2157409102827644\times 10^{-07}\,\xi^{1}_{1}\,\xi^{1}_{3}\,\eta^{2}_{1}\,
&  +5.8365071555190228\times 10^{-07}&\,\xi^{1}_{1}\,\xi^{1}_{3}\,\eta^{1}_{1}\,\eta^{1}_{2}\,\cr
&  -4.6671904570366227\times 10^{-06}\,\xi^{1}_{1}\,\xi^{1}_{3}\,\eta^{1}_{1}\,\eta^{1}_{3}\,
&  -8.0739065076924997\times 10^{-07}&\,\xi^{1}_{1}\,\xi^{1}_{3}\,\eta^{2}_{2}\,\cr
&  +5.4429327654203341\times 10^{-08}\,\xi^{1}_{1}\,\xi^{1}_{3}\,\eta^{1}_{2}\,\eta^{1}_{3}\,
&  +2.6230652324380928\times 10^{-05}&\,\xi^{1}_{1}\,\xi^{1}_{3}\,\eta^{2}_{3}\,\cr
&  -4.8323841400859345\times 10^{-03}\,\xi^{4}_{2}\,
&  +2.9298658121783215\times 10^{-05}&\,\xi^{3}_{2}\,\xi^{1}_{3}\,\cr
&  -1.3020117317952433\times 10^{-04}\,\xi^{2}_{2}\,\xi^{2}_{3}\,
&  -1.5832006427474452\times 10^{-03}&\,\xi^{2}_{2}\,\eta^{2}_{1}\,\cr
&  +7.6047082339419534\times 10^{-03}\,\xi^{2}_{2}\,\eta^{1}_{1}\,\eta^{1}_{2}\,
&  -8.0739065076924796\times 10^{-07}&\,\xi^{2}_{2}\,\eta^{1}_{1}\,\eta^{1}_{3}\,\cr
&  -9.6647220999081795\times 10^{-03}\,\xi^{2}_{2}\,\eta^{2}_{2}\,
&  +2.9299286278904711\times 10^{-05}&\,\xi^{2}_{2}\,\eta^{1}_{2}\,\eta^{1}_{3}\,\cr
&  -7.7905487286026464\times 10^{-05}\,\xi^{2}_{2}\,\eta^{2}_{3}\,
&  +1.9476359726545943\times 10^{-04}&\,\xi^{1}_{2}\,\xi^{3}_{3}\,\cr
&  +3.0033462029049320\times 10^{-07}\,\xi^{1}_{2}\,\xi^{1}_{3}\,\eta^{2}_{1}\,
&  -1.6360484568891460\times 10^{-06}&\,\xi^{1}_{2}\,\xi^{1}_{3}\,\eta^{1}_{1}\,\eta^{1}_{2}\,\cr
&  +5.4429327654202779\times 10^{-08}\,\xi^{1}_{2}\,\xi^{1}_{3}\,\eta^{1}_{1}\,\eta^{1}_{3}\,
&  +2.9299286278906453\times 10^{-05}&\,\xi^{1}_{2}\,\xi^{1}_{3}\,\eta^{2}_{2}\,\cr
&  -1.0427780546265602\times 10^{-04}\,\xi^{1}_{2}\,\xi^{1}_{3}\,\eta^{1}_{2}\,\eta^{1}_{3}\,
&  +1.9476665827159131\times 10^{-04}&\,\xi^{1}_{2}\,\xi^{1}_{3}\,\eta^{2}_{3}\,\cr
&  -2.0277494194124600\times 10^{-04}\,\xi^{4}_{3}\,
&  -7.4173186653203601\times 10^{-06}&\,\xi^{2}_{3}\,\eta^{2}_{1}\,\cr
&  +2.2482538047243565\times 10^{-07}\,\xi^{2}_{3}\,\eta^{1}_{1}\,\eta^{1}_{2}\,
&  +2.6230652324380681\times 10^{-05}&\,\xi^{2}_{3}\,\eta^{1}_{1}\,\eta^{1}_{3}\,\cr
&  -7.7905487286035477\times 10^{-05}\,\xi^{2}_{3}\,\eta^{2}_{2}\,
&  +1.9476665827159768\times 10^{-04}&\,\xi^{2}_{3}\,\eta^{1}_{2}\,\eta^{1}_{3}\,\cr
&  -4.0555535091919988\times 10^{-04}\,\xi^{2}_{3}\,\eta^{2}_{3}\,
&  -1.0838003720922736\times 10^{-04}&\,\eta^{4}_{1}\,\cr
&  +1.2014175808584629\times 10^{-03}\,\eta^{3}_{1}\,\eta^{1}_{2}\,
&  +6.2045352476790196\times 10^{-07}&\,\eta^{3}_{1}\,\eta^{1}_{3}\,\cr
&  -4.5563232782075760\times 10^{-03}\,\eta^{2}_{1}\,\eta^{2}_{2}\,
&  +8.8406443127175704\times 10^{-07}&\,\eta^{2}_{1}\,\eta^{1}_{2}\,\eta^{1}_{3}\,\cr
&  -9.7678628300066206\times 10^{-06}\,\eta^{2}_{1}\,\eta^{2}_{3}\,
&  +7.6046689202847939\times 10^{-03}&\,\eta^{1}_{1}\,\eta^{3}_{2}\,\cr
&  -2.4429460187142612\times 10^{-06}\,\eta^{1}_{1}\,\eta^{2}_{2}\,\eta^{1}_{3}\,
&  +3.9912387029285359\times 10^{-07}&\,\eta^{1}_{1}\,\eta^{1}_{2}\,\eta^{2}_{3}\,\cr
&  +2.6233130055604931\times 10^{-05}\,\eta^{1}_{1}\,\eta^{3}_{3}\,
&  -4.8323841400860802\times 10^{-03}&\,\eta^{4}_{2}\,\cr
&  +2.9298658121781443\times 10^{-05}\,\eta^{3}_{2}\,\eta^{1}_{3}\,
&  -1.3020117317952618\times 10^{-04}&\,\eta^{2}_{2}\,\eta^{2}_{3}\,\cr
&  +1.9476359726546422\times 10^{-04}\,\eta^{1}_{2}\,\eta^{3}_{3}\,
&  -2.0277494194122486\times 10^{-04}&\,\eta^{4}_{3}
\cr
&+o\big(\|(\vet{\csi},\vet{\eta})\|^4\big)\ .
\cr
\end{align*}

\bibliography{sec_SJSU_2D}
\bibliographystyle{model1b-num-names}

\end{document}